\documentclass[twocolumn,prd,twocolumn,superscriptaddress,nofootinbib]{revtex4-1}
\usepackage[latin9]{inputenc}
\usepackage{color}
\usepackage{verbatim}
\usepackage{units}
\usepackage{amsmath}
\usepackage{amssymb}
\usepackage{graphicx}
\PassOptionsToPackage{normalem}{ulem}
\usepackage{ulem}
\usepackage[unicode=true,pdfusetitle,
 bookmarks=true,bookmarksnumbered=false,bookmarksopen=false,
 breaklinks=false,pdfborder={0 0 1},backref=false,colorlinks=true]
 {hyperref}
\hypersetup{
 citecolor=blue}

\makeatletter

\@ifundefined{textcolor}{}
{%
  \definecolor{BLACK}{gray}{0}
  \definecolor{WHITE}{gray}{1}
  \definecolor{RED}{rgb}{1,0,0}
  \definecolor{GREEN}{rgb}{0,1,0}
  \definecolor{BLUE}{rgb}{0,0,1}
  \definecolor{CYAN}{cmyk}{1,0,0,0}
  \definecolor{MAGENTA}{cmyk}{0,1,0,0}
  \definecolor{YELLOW}{cmyk}{0,0,1,0}
}

\usepackage{bm}
\usepackage{braket}
\usepackage{cancel}

\makeatother

\begin{document}
\title{Constraints from a many-body method on spin-independent dark matter scattering off electrons
using data from germanium and xenon detectors}
\author{Mukesh K. Pandey}
\affiliation{Department of Physics, National Taiwan University, Taipei 10617, Taiwan}
\author{Lakhwinder~Singh}
\affiliation{Department of Physics, Central University of South Bihar, Gaya 824236, India}
\affiliation{Institute of Physics, Academia Sinica, Taipei 11529, Taiwan}
\author{Chih-Pan~Wu}
\affiliation{Department of Physics, National Taiwan University, Taipei 10617, Taiwan}
\author{Jiunn-Wei~Chen}
\email{jwc@phys.ntu.edu.tw}

\thanks{corresponding author}
\affiliation{Department of Physics, Center for Theoretical Physics, and Leung Center
for Cosmology and Particle Astrophysics, National Taiwan University,
Taipei 10617, Taiwan}
\affiliation{Center for Theoretical Physics, Massachusetts Institute of Technology,
Cambridge, MA 02139, USA}
\author{Hsin-Chang~Chi}
\affiliation{Department of Physics, National Dong Hwa University, Shoufeng, Hualien
97401, Taiwan}
\author{Chung-Chun Hsieh}
\affiliation{Department of Physics, National Taiwan University, Taipei 10617, Taiwan}
\author{C.-P.~Liu}
\email{cpliu@mail.ndhu.edu.tw}

\thanks{corresponding author}
\affiliation{Department of Physics, National Dong Hwa University, Shoufeng, Hualien
97401, Taiwan}
\author{Henry~T.~Wong}
\affiliation{Institute of Physics, Academia Sinica, Taipei 11529, Taiwan}
\date{\today}
\begin{abstract}
Scattering of light dark matter (LDM) particles with atomic electrons
is studied in the context of effective field theory. Contact and long-range
interactions between dark matter and an electron are both considered.
A state-of-the-art many-body method is used to evaluate the spin-independent
atomic ionization cross sections of LDM-electron scattering, 
with an estimated error about $ 20\%$. 
New upper limits are derived on parameter space spanned by LDM mass and effective
coupling strengths using data from the CDMSlite, XENON10, XENON100,
and XENON1T experiments. 
Comparison with existing calculations shows the importance of atomic 
structure. Two aspects particularly important are relativistic effect for inner-shell ionization, and
final-state free electron wave function which sensitively depends on the underlying atomic approaches.
\end{abstract}
\maketitle

\section{Introduction.}

Astronomical and cosmological observations not only provide evidences
of dark matter (DM) but also point out its properties such as nonrelativistic,
non-baryonic, stable with respect to cosmological time scale, and
interacting weakly, if any, with the standard model (SM) particles.
Its non-gravitational interactions with normal matter are still unknown.
A generic class of cold dark matter candidates, the so-called Weakly-Interacting
Massive Particles (WIMPs), receive most attention, as they lead to
predictions of DM's relic abundance comparable to the measured value
and have coupling strengths of weak interaction scales to SM particles,
which can be experimentally tested. Also, the existence of such particles
are predicted in many extensions of the SM (see, e.g., Refs.~\cite{Tanabashi:2018oca}
for review). Recently there has been remarkable progress made in direct
WIMP searches, thanks to novel innovations in detector technologies
and increment of detector size and exposure time. As a result, a substantial
portion of the favored WIMP parameter space has now been ruled out.
For example, the  stringent bounds
on the spin-independent WIMP-nucleon cross section are currently set
by the XENON1T~\cite{Aprile:2018dbl}: $4.1\times10^{-47}\textrm{cm}^{2}$
at $30\,\textrm{GeV}$ dark matter mass,~\footnote{We use the natural units $\hbar=c=1$. },
PandaX-II~\cite{PandaX-II:2017}: $8.6\times10^{-47}\textrm{cm}^{2}$
at $40\,\textrm{GeV}$,
and DarkSide-50~\cite{Agnes:2018ves}: $1\times10^{-41}\,\textrm{cm}^{2}$
at $1.8\,\textrm{GeV}$, respectively.

In spite of tremendous efforts in experiment, no concrete evidence
of WIMPs has been found to date, directly or indirectly. This motivates
searches of DM particles with masses lighter than generic WIMPs, i.e.,
$\lesssim10\,\textrm{GeV}/c^{2}$. Theoretically, such light dark
matter (LDM) candidates arise in many well-motivated models, and to
account for the relic DM abundance, there are mechanisms suggesting
LDM interacts with SM particles through light or heavy mediators with
coupling strengths smaller than the weak scale~(see Ref.~\cite{Essig:2013lka}
for review). Moreover, annihilations or decays of LDM candidates are
possible sources of the anomalous 511 keV~\cite{Knodlseder:2005yq,Finkbeiner:2007kk}
and 3.5 keV~\cite{Bulbul:2014sua,Boyarsky:2014jta} emission lines
recently found in the sky. Consequently, new ideas to search for LDM
flourish and have good discovery potential (see Ref.~\cite{Battaglieri:2017aum}
for general survey).

The energy transfer of an incident DM particle to a target particle
depends on the reduced mass of the system. Current low-threshold experiments sensitive to sub-keV nuclear recoil, such as CoGENT~\cite{Aalseth:2011wp,Aalseth:2012if,Aalseth:2014eft} and CDEX~\cite{Yang:2017yaw,Jiang:2018pic,Yang:2019lao}, can only search for DM particles with masses as low as a few GeV through DM-nucleus interactions. Various other experiments such as CRESST-II~\cite{CRESST:2016}, DAMIC~\cite{DAMIC:2012}, NEWS-G~\cite{NEWS-G:2017}, PICO~\cite{PICO:2017}, SENSEI~\cite{SENSEI:2017,Crisler:2018gci,Abramoff:2019dfb}, and SuperCDMS~\cite{SuperCDMS:2016,superCDMS:PRD2018,SuperCDMS:2018},
are pursuing intensive research programs towards lower detector threshold
and therefore lower mass.

For energy deposition in the sub-keV region, electron recoil becomes
an important subject, no matter being taken as a signal or background,
because LDM particles transfer their kinetic energy more efficiently
to target electrons than nuclei. Furthermore, electron recoil signals
can be used to directly constrain LDM-electron interactions; this
complements the study of LDM-nucleon interactions through nuclear
recoil and extends a direct detector's scientific reach. Constraints
of LDM-electron scattering by direct detection experiments emerged
recently, e.g., DAMA/LIBRA~\cite{Roberts:2015lga,Roberts:2016xfw},
DarkSide-50~\cite{DarkSide:2018}, SuperCDMS~\cite{superCDMS:PRD2018},
XENON10~\cite{Essig:PRL2012,EssigPRD:2017}, XENON100~\cite{XENON100:2015,EssigPRD:2017},
and XENON1T~\cite{Aprile:2019xxb}; and much improvement
will certainly be expected in next-generation sub-keV detectors.

While electron recoil at sub-keV energies opens a new, exciting window
for LDM searches, the scattering processes of LDM particles in detectors
pose a fundamental theoretical challenge: The typical energy and momentum
of a bound electron is on the order of $Z_{\textrm{eff}}\,m_{e}\alpha$
and $Z_{\textrm{eff}}^{2}\,m_{e}\alpha^{2}/2$, respectively, where
$Z_{\textrm{eff}}$ is the effective nuclear charge felt by an electron
of mass $m_{e}$ in a certain shell and $\alpha$ is the fine structure
constant with $m_{e}\alpha\approx3.7\,\textrm{keV}$. Consequently
a sub-keV scattering event strongly overlaps with the atomic scales.
This implies a reliable calculation of LDM-electron scattering cross
section, which is needed for data analysis, should properly take into
account not only the bound nature of atomic electrons but also the
electron-electron correlation.

In this work, we applied a state-of-the-art many-body method to evaluate
the atomic ionization cross sections of germanium (Ge) and xenon (Xe)
by spin-independent LDM-electron scattering. New upper limits on parameter
space spanned by the effective coupling strengths and mass of LDM
are derived with data from CDMSlite~\cite{superCDMS:PRD2018}, XENON10~\cite{Xenon10:2011},
XENON100~\cite{Xenon100:2016}, and XENON1T~\cite{Aprile:2019xxb}.
The results are also compared with existing calculations.

\section{Formalism. }

A general framework for dark matter interaction with normal matter
has recently been developed using effective field theory (EFT). This
framework accommodates scalar, fermionic, and vector nonrelativistic
(NR) DM particles interacting with NR nucleons via scalar and vector
mediators~\cite{EFT:2010}. All leading-order and next-to-leading-order
operators in the effective DM-nucleon interaction are identified~\cite{NLO-EFT:2012}.
The DM-electron interaction can be formulated similarly with the electron
being treated relativistically, as it is essential for atomic structure
of Ge and Xe~\cite{Chen:2015pha}. At leading order (LO), the spin-independent
(SI) part is parametrized by two terms: 
\begin{align}
\mathcal{L}_{\textrm{SI}}^{(\textrm{LO})}= & c_{1}(\chi^{\dagger}\chi)(e^{\dagger}e)+d_{1}\frac{1}{q^{2}}(\chi^{\dagger}\chi)(e^{\dagger}e)\,,\label{eq:L_SI}
\end{align}
where $\chi$ and $e$ denote DM and electron fields, respectively,
and $q=|\vec{q}|$ is the magnitude of 3-momentum transfer, which
can be determined by the NR DM particle's energy transfer $T$ and
scattering angle $\theta$. 
The low-energy constants $c_{1}$ and $d_{1}$ characterize the strengths
of the short-range (for heavy mediators) and long-range (for light
mediators) interactions, respectively. While the masses of the mediators
can vary in broad ranges, it is customary to consider the two extremes:
the extremely massive and the massless, which give rise to the contact
(or zero-range) and the (infinitely) long-range interactions, respectively.

The main scattering process that yields electron recoil is atomic
ionization: 
\begin{equation}
\chi+\textrm{A}\rightarrow\chi+\textrm{A}^{+}+e^{-}\,,\label{eq:AI}
\end{equation}
and the energy deposition by DM is reconstructed by subsequent secondary
particles, such as photons and more ionized electrons, recorded in
a detector. The differential DM-atom ionization cross section in the
laboratory frame through the LO, SI DM-electron interaction is derived
in Ref.~\cite{Chen:2015pha}
\begin{equation}
\frac{d\sigma}{dT}=\frac{1}{2\pi v_{\chi}^{2}}\int dq\, q\left[\left|c_{1}+\frac{d_{1}}{q^{2}}\right|^{2}\right]R(T,q),\label{eq:dcs-LO}
\end{equation}
where $m_{\chi}$ and  $v_{\chi}$ are the mass and velocity  of the DM particle, respectively.

The full information of how the detector atom responds to the incident
DM particle is encoded in the response function 
\begin{align}
R(T,\theta)= & 
\sum_{i=1}^{Z}\int d^{3}p_{i}\,|\braket{\textrm{A}^{+},e^{-}||e^{i\frac{\mu}{m_{e}}\vec{q}.\vec{r}_{i}}||\textrm{A}}|^{2}\nonumber \\
 & \times\delta(T-E_{B_{i}}-\frac{\vec{q}^{2}}{2M}-\frac{\vec{p_{i}}^{2}}{2\mu})\,,\label{eq:response}
\end{align}
where $\ket{\textrm{A}}$ and $\ket{\textrm{A}^{+},e^{-}}$ denote
the many-body initial (bound) and final (ionized) state; $M$ and
$\mu$ the total and reduced mass of the ion plus free electron system,
respectively, with $\mu\approx m_{e}$. The summation is over all
electrons, and the $i$th electron has its binding energy $E_{B_{i}}$,
relative coordinate $\vec{r}_{i}$, and relative momentum $\vec{p}_{i}$.
The Dirac delta function imposes energy conservation and constrains
the kinematics of the ejected electron, whose energy is NR in the
kinematic range of our study but wave function still in the fully
relativistic form.

Evaluation of $R(T,q)$ is non-trivial.
In this work, we use a procedure benchmarked by photoabsorption to a few percent accuracy
based on an \textit{ab
inito} method, the (multi-configuration) relativistic random phase
approximation, (MC)RRPA~\cite{Johnson:1979wr,Johnson:1979ch,Huang:1981wj,Huang:1982re,Huang:1995cc}.
Details of how the theory was applied to the responses of Ge and Xe
detectors in cases of neutrino scattering are documented in Refs.~\cite{Chen:2013lba,Chen:2014dsa,Chen:2014ypv,Chen:2016eab};
here we only give a brief outline and focus on the points that are
new in the case of LDM scattering.

\begin{figure*}
	\begin{tabular}{cc}
		\includegraphics[width=8cm]{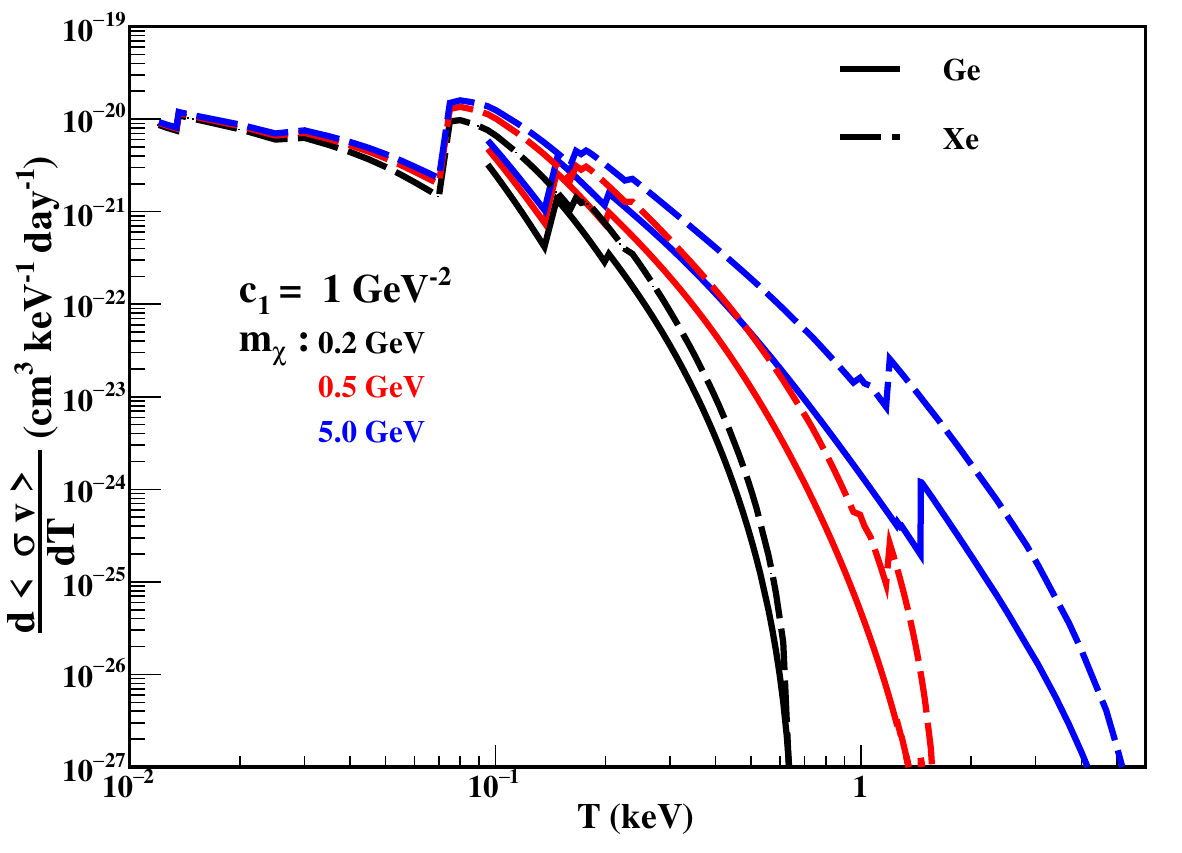} & \includegraphics[width=8cm]{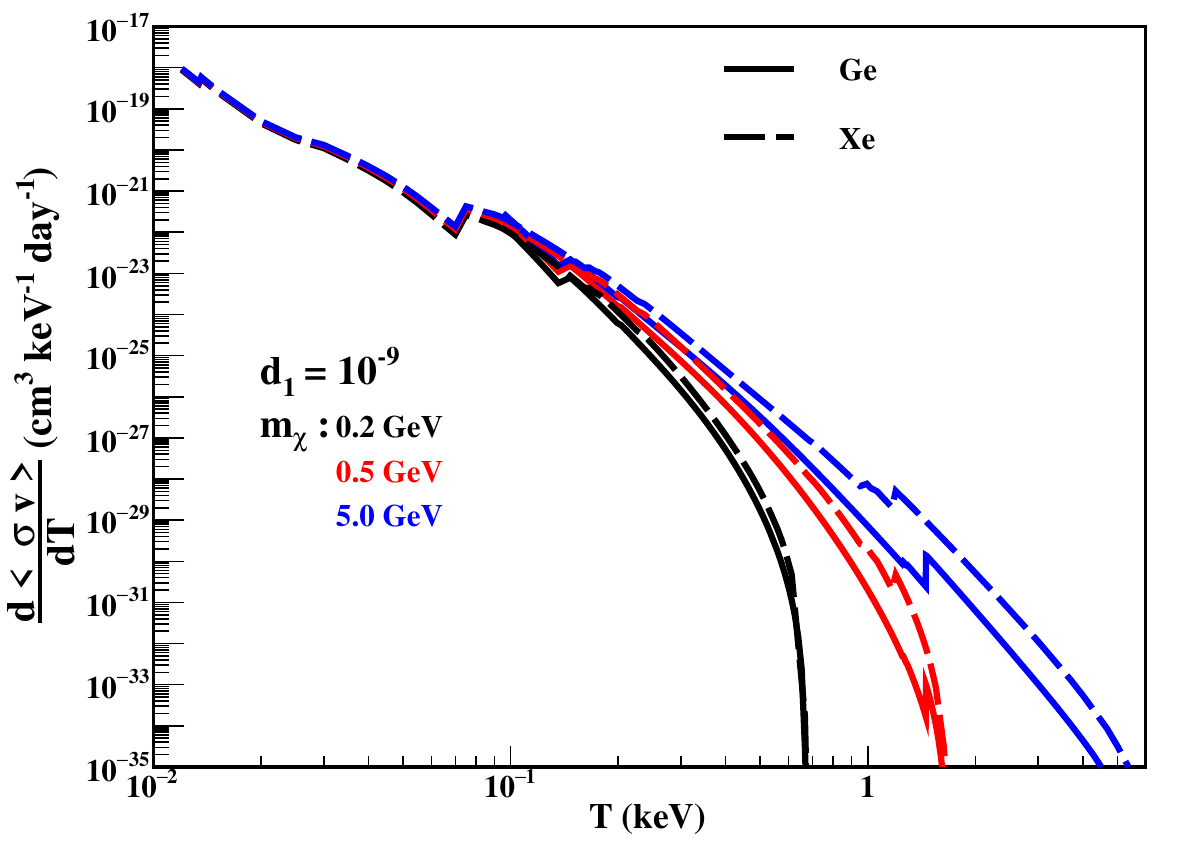}\tabularnewline
	\end{tabular}\caption{
		Averaged velocity-weighted differential
		cross sections for ionization of Ge and Xe atoms by LDM of various
		masses with the effective spin independent short-range (left) and long-range (right)
		interactions where $c_{1}=1/\textrm{GeV}^{-2}$ and $d_{1}=10^{-9}$.}
	\label{fig::dcs_ave}
\end{figure*}

First, the ground-state atomic wave functions are calculated by (MC)
Dirac-Fock (DF) theory. The multi-configuration feature is needed
for open-shell atoms like Ge, but not for noble gas atoms like Xe.
Quality of the initial wave function is benchmarked by the ionization
energies of all atomic shells, which can be determined by edges in
photoabsorption data.

Second, (MC)RRPA is applied to calculate the transition matrix elements
of photoionization. Quality of the final state wave function is benchmarked
by how good the calculated photoionization cross section is compared
with experiment data. For Ge and Xe, the atomic numbers $Z=32$ and
$54$ are not small, so relativistic corrections to inner-shell electrons
or at large 3-momentum transfer are not negligible (a detailed study
can be found in Refs.~\cite{Roberts:2015lga,Roberts:2016xfw}).
Furthermore, the residual electron correlation is important for excited
states, as a result, its (partial) inclusion by RPA makes our calculated
photoionization cross sections of Ge and Xe in excellent agreement
with experiments. The only exception is $T<80\,\textrm{eV}$ for Ge,
where the crystal structure of outer-shell electrons in Ge semiconductor
can not be described by our pure atomic calculations~\cite{Chen:2013lba,Chen:2014ypv}.

Taking the well-benchmarked initial and final state wave functions, response functions for DM-atom scattering, Eq.~(\ref{eq:response}), can in principle be computed in a similar procedure as we previously did for neutrino-atom scattering. However, the high 3-momentum transfer associated with the DM-atom scattering, $q\gtrsim m_{e}\alpha$, dramatically slows down the (MC)RRPA computation. Therefore, we resorted to a conventional frozen-core approximation (FCA, see, e.g., Refs.~\cite{Slater_1951hf,Sampson:1989df}), so that the final-state continuum wave function of the ionized electron can be efficiently computed with an electrostatic mean field determined from the ionic state prescribed by (MC)DF. The details of our FCA scheme and its benchmark against (MC)RRPA are given in Appendix A.  Except when energy transfer is close to ionization edges, the FCA results generally agree with (MC)RRPA within $20\%$.

At a direct detector, the measured event rate is 
\begin{eqnarray}
\frac{d\mathcal{R}}{dT}=\frac{\rho_{\chi}\,N_{T}}{m_{\chi}}\frac{d\braket{\sigma v_{\chi}}}{dT}\,,\label{eq:count-rate}
\end{eqnarray}
where $\rho_{\chi}=0.4$ GeV/cm$^{3}$ is the local DM density~\cite{DM-density:2009},
and $N_{T}$ is the number of target atoms. The averaged velocity-weighted
differential cross section 
\begin{equation}
\frac{d\braket{\sigma v_{\chi}}}{dT}=\int_{v_{\min}}^{{v_{\textrm{max}}}}d^{3}v_{\chi}f(\vec{v}_{\chi})v_{\chi}\frac{d\sigma}{dT}\,,\label{eq:vchi_ave}
\end{equation}
is folded to the conventional Maxwell-Boltzmann velocity distribution
$f(\vec{v}_{\chi})$ \cite{Lewin:1995rx}, with
escape velocity $v_{\textrm{esc}}=544\,\textrm{km/s}$~\cite{Smith:2006ym},
circular velocity $v_{0}=220\,\textrm{km/s}$, and averaged Earth
relative velocity $v_{E}=232\,\textrm{km/s}$. The maximum DM velocity
seen from the Earth is $v_{\textrm{max}}=v_{\textrm{esc}}+v_{E}$,
and the minimum $v_{\min}=\sqrt{2T/m_{\chi}}$ is to guarantee enough
kinetic energy. This velocity average is time-consuming because $d\sigma/dT$ needs
to be computed on every grid point of $v_{\chi}$. As commonly seen
in literature, e.g., Refs.~\cite{Kopp:2009et,Essig:2011nj}, the procedure is simplified by an interchange of integration order: first on $v_{\chi}$ of Eq.~(\ref{eq:vchi_ave}) with $\frac{1}{v^2_{\chi}}$ of Eq.~(\ref{eq:dcs-LO}) ~\cite{Lewin:1995rx,Savage:2008er},  then the remaining $q$ integration. The details is given in Appendix~B\ref{sec:eta}, and we numerically checked that the two average schemes agree very well for
all kinematics considered in this work.

In Fig.~\ref{fig::dcs_ave}, we show some results of $d\braket{\sigma v_{\chi}}/dT$
for Ge and Xe targets with selected LDM masses. There are several
noticeable features. First, the sharp edges correspond to ionization
thresholds of specific atomic shells. They clearly indicate the effect
of atomic structure, and the peak values sensitively depend on atomic
calculations. If direct DM detectors have good enough energy resolution,
these peaks can serve as powerful statistical hot spots. Second, away
from these edges, the comparison between Ge and Xe cases do point
out that the latter has a larger cross section, but the enhancement
is not as strong as $Z^{2}$ for coherent scattering nor $Z$ for
incoherent sum of free electrons. In other words, a heavier target
atom does not enjoy much advantage in constraining the SI DM-electron
interaction, in opposition to the SI DM-nucleon case. Third, the long-range
interaction has a larger inverse energy dependence than the short-range
one. As a result, lowering threshold can effectively boost a detector's
sensitivity to the long-range DM-electron interaction.

\section{Results and Discussions}

\begin{figure*}
	\begin{tabular}{cc}
		\includegraphics[width=8.5cm]{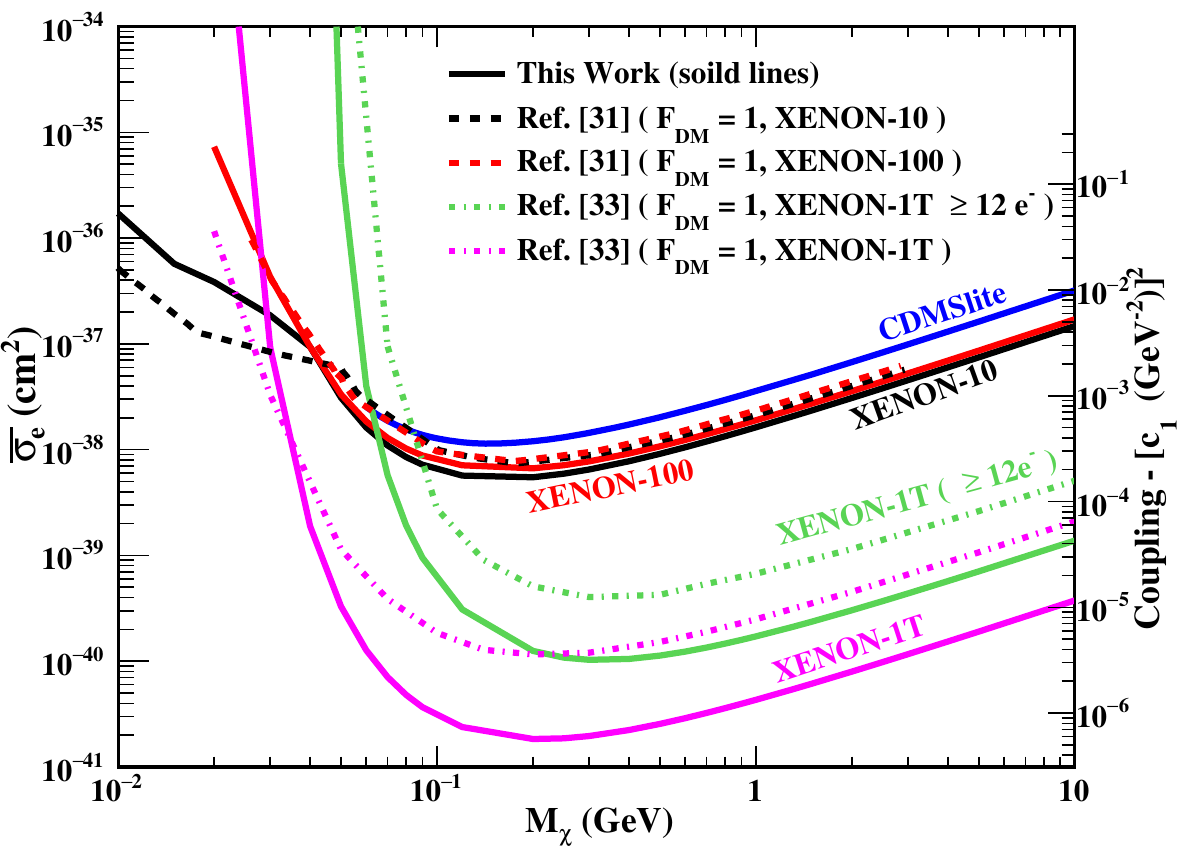}  & \includegraphics[width=8.5cm]{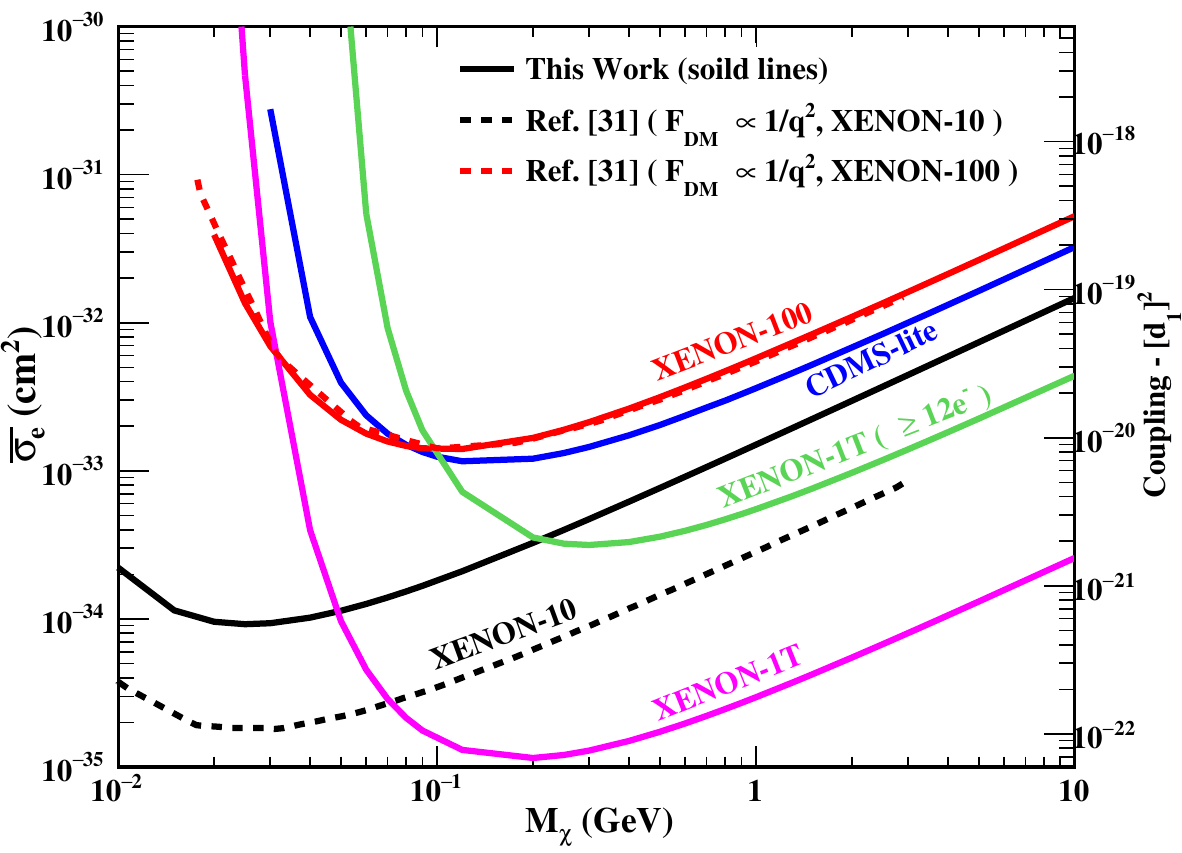} \tabularnewline
	\end{tabular}\caption{
		Exclusion limits at 90\% C.L. on
		the spin-independent short-(left) and long-(right) range LDM-electron
		interactions as functions of $m_{\chi}$ derived from CDMSlite (blue)~\cite{superCDMS:PRD2018},
		XENON10 (black)~\cite{Xenon10:2011}, XENON100 (red)~\cite{Xenon100:2016}, 
		and XENON1T (magenta and green)~\cite{Aprile:2019xxb} data. Superimposed
		are constraints from Ref.~\cite{EssigPRD:2017} with XENON10 (black-dotted) and  XENON100 (red-dotted), and from Ref.~\cite{Aprile:2019xxb} with XENON1T (magenta-dotted and green-dotted, short-range only), in which the choices of $F_{DM}=1$ (left)
		and $F_{DM}=1/q^{2}$ (right) correspond to $c_{1}$- and $d_{1}$-type
		interactions, respectively, in this work.}
	\label{fig:ex-plot} 
\end{figure*}

The CDMSlite experiment with Ge-crystals as target has recently demonstrated
the novel mechanism of bolometric amplification~\cite{Agnese:2013jaa}
and achieved low ionization threshold making it sensitive to LDM searches.
A data set of 70.1 kg-day exposure~\cite{superCDMS:PRD2018} and
threshold of 80 eV is adopted for this analysis. The combined trigger
and pulse-shape analysis efficiency is more than 80\%. 
The fiducial-volume cut significantly reduced the background and the total combined efficiency including  fiducial-volume is adopted from Fig.~4. of Ref.~\cite{superCDMS:PRD2018}.
Limits on LDM-electron scattering are derived without background subtraction with optimum
interval method~\cite{Yellin:2008}. The derived 90\% C.L. limits
for both short- and long-range coupling are depicted in Fig.~\ref{fig:ex-plot}.

Dual-phase liquid Xe detectors have demonstrated the sensitivity to
ionization of a single electron with their ``S2-only'' signals~\cite{XENON10:Asrto}.
Constraints have been placed in Refs.~\cite{Essig:PRL2012,EssigPRD:2017}
with XENON10~\cite{Xenon10:2011} , XENON100~\cite{Xenon100:2016},
and XENON1T\cite{Aprile:2019xxb} data on LDM-electron scattering
using an alternative theoretical framework with different treatment
to the atomic physics from this work.

Efficiency-corrected data of XENON10~\cite{Xenon10:2011,Essig:PRL2012}
with 15 kg-day exposure, XENON100~\cite{Xenon100:2016} with 30
kg-year exposure, and XENON1T\cite{Aprile:2019xxb} with 1 ton-year exposure are extracted
from the literature. We follow the same procedure of Refs.~\cite{Essig:PRL2012,EssigPRD:2017}
to convert energy transfer $T$ first to the number of secondary electrons,
$n_{e}$, and then to the photoelectron (PE) yield. Under a conservative
assumption that all observed events are from potential LDM-electron
scattering, upper limits at 90\% C.L. on both short- and long-range
interactions are derived and displayed in Fig.~\ref{fig:ex-plot}.
The electron recoil charge yield ($Q_y$) cutoff can change the exclusion region. For the analysis of XENON1T data~\cite{Aprile:2019xxb}, we present both results with and without a cutoff at $12$ produced electrons. For the latter, events of smaller ionized electrons enter through Gaussian smearing of PEs.

Comparing the various exclusion curves in Fig.~\ref{fig:ex-plot} 
there are several important observations to note. First, the lowest
reach of a direct search experiment in LDM mass is determined by its
energy threshold. According to what we set for CDMSlite, XENON10, XENON100,
and XENON1T: $80$, $13.8$, $56$, and
$186\,\textrm{eV}$, the lightest DM masses can be probed are $\sim 30$,
$10$, $20$, and $50\,(25\,\textrm{without } Q_y \textrm{ cutoff})\,\textrm{MeV}$, respectively.

The exclusion limits on DM-electron interaction strengths depend on
several factors: experimentally, detector species, energy resolution,
background, and exposure mass-time~\cite{Undagoitia:2015gya}; and theoretically, the DM-electron interaction type and the atomic structure. For the contact interaction, XENON1T yields the best limit when $m_{\chi} \gtrsim 50\,(25\,\textrm{without } Q_y \textrm{ cutoff})\,\textrm{MeV}$ for its overwhelmingly-large exposure mass-time. However, in the lower-mass region, it is the extremely-low threshold of XENON10 that makes it more competitive, despite a much smaller exposure mass-time. 
The exclusion limit on the long-range interaction is more subtle, because the differential cross section has a sharper energy dependence and tends to weight more at low $T$. This explains why XENON10's constraint is better than others (less so when XENON1T has no $Q_y$ cutoff) in the plot. Among other three experiments, the dominance of XENON1T is shrinking for the same reason. More surprisingly, the finer energy resolution and lower background of CDMSlite achieves a better constraint than XENON100 when $m_{\chi} \gtrsim 80\,\textrm{MeV}$, despite the latter has a big exposure advantage.

\begin{figure*}
	\begin{tabular}{cc}
		\includegraphics[width=8.5cm]{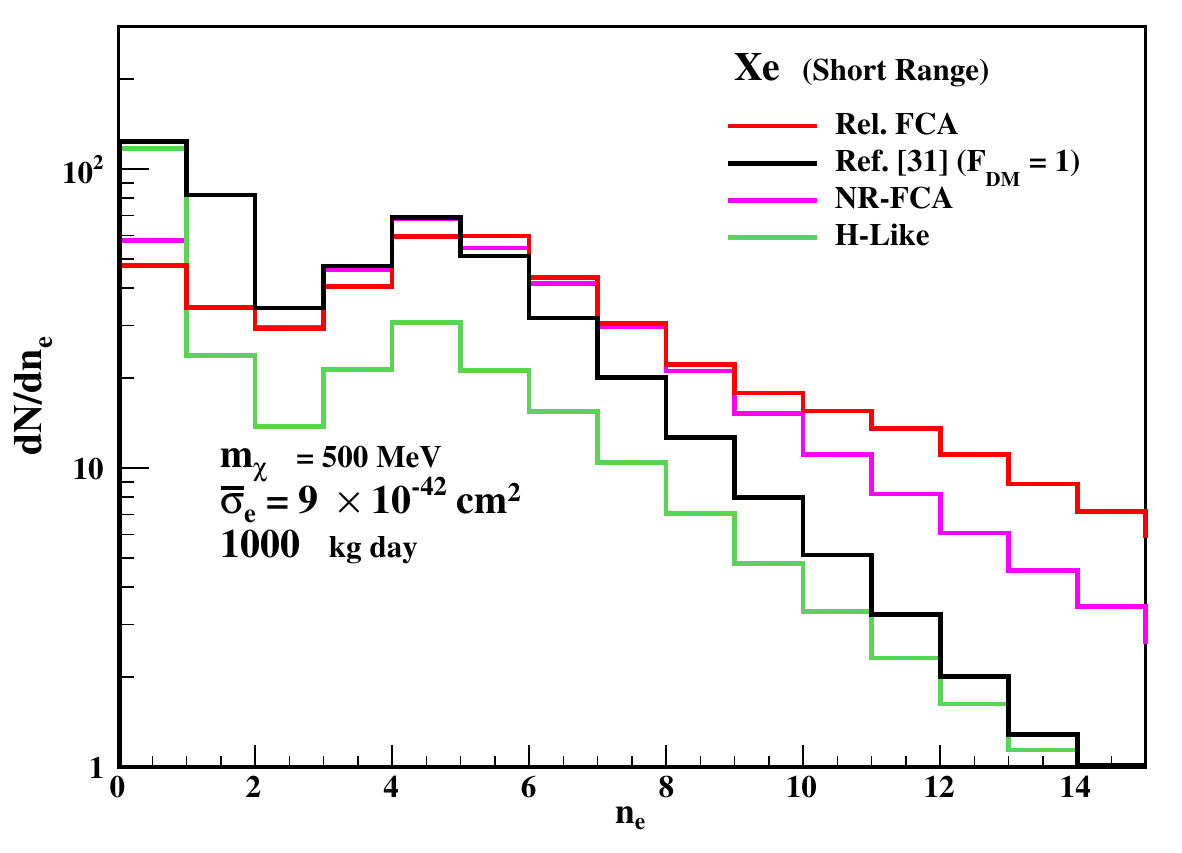}  & \includegraphics[width=8.5cm]{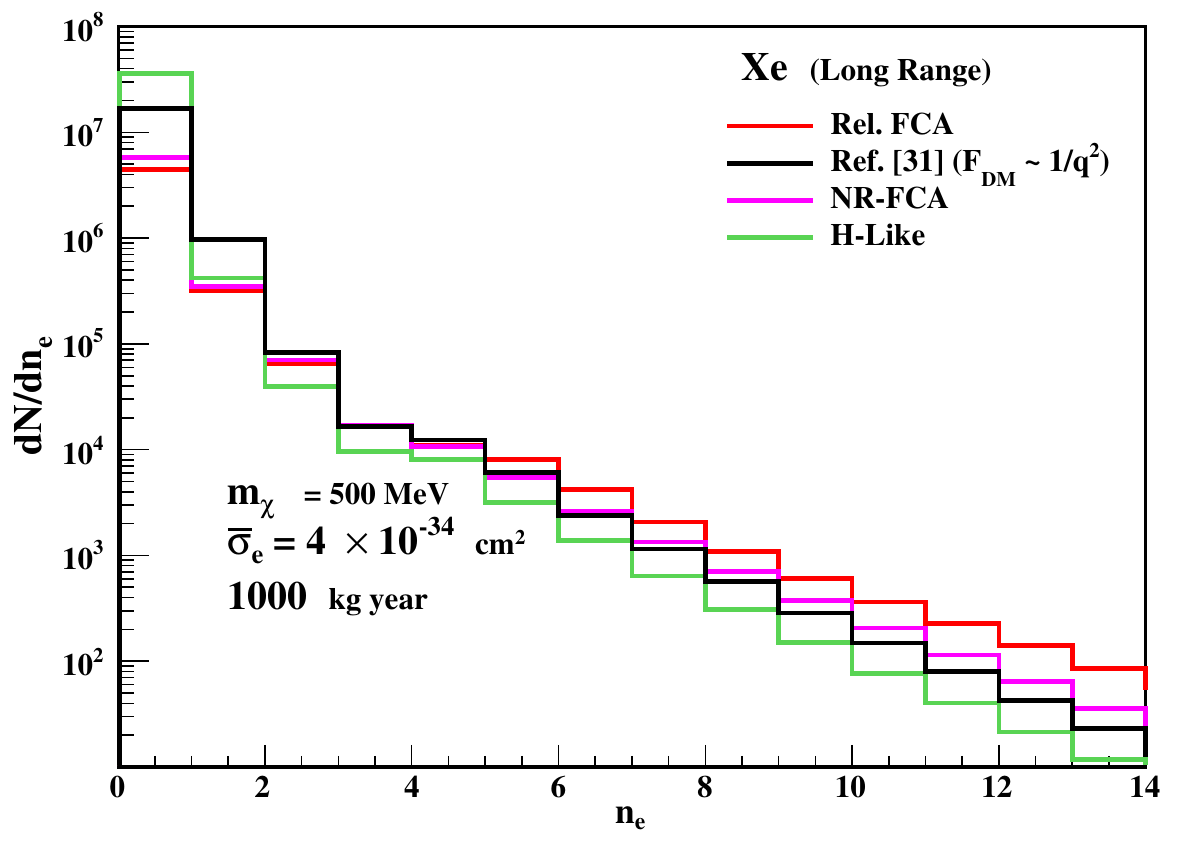} \tabularnewline
	\end{tabular}\caption{
		Comparisons of expected event
		numbers as a function of ionized electron number derived in this work
		(relativistic FCA, red), non-relativistic FCA (blue), hydrogen-like approximation (green), and from Ref.~\cite{EssigPRD:2017} (black) for
		Xe detectors with 1000 kg-year exposure, assuming DM mass $m_{\chi}=500\,\textrm{MeV}$,
		and DM-electron interaction strengths (left) $c_{1}=5.28\times10^{-4}\,\textrm{GeV}^{-2}$
		and (right) $d_{1}=4.89\times10^{-11}$ (equivalent to $\sigma_{e}=9\times10^{-42}\,\textrm{cm}^{2}$
		and $\bar{\sigma}_{e}=4\times10^{-34}\,\textrm{cm}^{2}$, respectively,
		in Ref.~\cite{EssigPRD:2017}).}
	\label{fig:comp} 
\end{figure*}

In Fig.~\ref{fig:ex-plot}, the exclusion limits derived in Refs.~\cite{EssigPRD:2017,Aprile:2019xxb},
using the same xenon data sets, are compared.~\footnote{The conversion $\sigma_{e}=c_{1}^{2}\mu_{\chi e}^{2}/\pi$ and $\bar{\sigma}_{e}=d_{1}^{2}\mu_{\chi e}^{2}/\left(\pi(m_{e}\alpha)^{4}\right)$
with $\mu_{\chi e}=m_{\chi}m_{e}/(m_{\chi}+m_{e})$ is used.} The differences in the overall exclusion curves are obvious and most
likely of theoretical origins. In Fig.~\ref{fig:comp}, we use a
sample case to illustrate the difference in predicted event numbers
as a function of $n_{e}$. For both types of interactions, our results
are comparatively smaller at small $n_{e}$ but bigger at large $n_{e}$.
This provides a qualitative explanation for the overall differences
observed in the exclusion curves: The larger the DM mass $m_{\chi}$,
the larger its kinetic energy and hence the increasing chance of higher
energy scattering that produces more $n_{e}$. Therefore, our calculations
yield tighter constraints on $c_{1}$ for heavier DM particles, but
looser for lighter DM particles. As for the long-range interaction,
the low-energy cross section is so dominant that the derivation of
exclusion limit with the XENON10 data is dictated by the one-electron event, i.e., the first
bin. Consequently, the larger event number predicted in Ref.~\cite{EssigPRD:2017} leads to a better 
constraint on $d_{1}$ by a similar size. As for XENON100, its higher energy threshold cut out 
most sensitivity to low $n_e$ events, so both exclusion curves become similar.  
While there is no exclusion from Ref.~\cite{Aprile:2019xxb} to compare, we note that the analysis
without $Q_y$ cutoff does improve the constraint substantially.

To further trace the main origin of this discrepancy,
we performed two additional sets of calculations: (I) The non-relativistic
frozen core approximation (NRFCA): In this approach, the electrostatic
mean field used to calculate the continuum final states are based
on the same Roothaan-Hartree-Fock (RHF) orbital wave functions~\cite{Clementi:1974yvo,Bunge:1993jsz} as Refs.~\cite{Essig:PRL2012,EssigPRD:2017}. (II) The hydrogen-like approximation (HLA) 
as in Refs.~\cite{Essig:2011nj,Agnes:2018ves,Catena:2019gfa}: In this
approach, a continuum final state is simply the Coulomb wave function
with an effective nuclear charge determined from the binding energy
of the atomic shell from which the electron is ionized. In Appendix
C, we discuss and compare these atomic approaches in more detail.

As the comparison of FCA vs. NRFCA shows in Fig.~\ref{fig:comp}, for $n_{e} \ge 6$, the relativistic effect 
becomes increasingly important, because the inner shell (in this case, $4d$) ionization dominates the cross section. This is consistent with what has been reported in Refs.~\cite{Roberts:2015lga,Roberts:2016xfw}. 
However, this effect is not enough to explain the discrepancy with Ref.~\cite{EssigPRD:2017}. The difference between the NRFCA and Ref.~\cite{EssigPRD:2017} is most likely due to different formulations of the effective Coulomb potential felt by an ionized electron.  However, no further comment can be made as the detail is not explicitly given in Refs.~\cite{Essig:PRL2012,EssigPRD:2017}. On the other hand, we did find the results of Ref.~\cite{EssigPRD:2017} falls in between NRFCA and HLA, so perhaps is the reconstructed Coulomb potentials.

We note that a later relativistic calculation by Roberts et al.~\cite{Roberts:2019chv} is similar to our framework, but differs in the formulation of the frozen core potential. As a result, even though the agreement of both calculations are generally good, there is still difference due to atomic treatments.

\section{Summary}

In summary, we conclude the scattering cross section of sub-GeV dark matter off atoms depends sensitively on atomic structure. Two aspects particularly important are: the relativistic effect, which is sizable when inner-shell electrons are ionized, and the final-state wave functions on which electron correlation plays an important role. The atomic approached used in our work is fully relativistic, and as the benchmark with (MC)RRPA (a truly many-body approach) validates that our adopted frozen-core mean field is quite robust, and the theoretical uncertainty of our results is estimated to be about $20\%$.

\begin{acknowledgments}

This work is supported in part under contracts 104-2112-M-001-038-MY3,
104-2112-M-259-004-MY3, 105-2112-002-017-MY3, 107-2119-001-028-MY3,
and 107-2112-M-259-002 from the Ministry of Science and Technology,
and 2017-18/ECP-2 from the National Center of Theoretical Physics,
of Taiwan.

\end{acknowledgments}

\appendix

\section{A. Frozen Core Approximation~\label{sec:FCA}}

In this appendix, we give an outline of our formulation of the frozen
core potential, by which the final ionized electron wave function
is computed.

Starting from the self-consistent, relativistic mean field theory,
the (MC)DF routine yields a set of single particle orbitals which
take a 2-spinor form 
\begin{equation}
\phi_{a,m_{a}}(\vec{r})=\frac{1}{r}\left(\begin{array}{c}
g_{a}(r)\Omega_{\kappa_{a},m_{a}}(\hat{r})\\
if_{a}(r)\Omega_{-\kappa_{a},m_{a}}(\hat{r})
\end{array}\right)\,,\label{eq:rel_wf}
\end{equation}
where $a=(n_{a},\kappa_{a})$ is a collective label for the principle
($n_{a}$) and relativistic orbital ($\kappa_{a}$) quantum numbers
of the $a$th atomic shell; $g_{a}$ and $f_{a}$ the reduced radial
wave functions of the large and small components, respectively; and
$\Omega_{\kappa_{a},m_{a}}(\hat{r})$ the spin-angular function that
depends on $\kappa_{a}$, the solid angle $\hat{r},$ and the total
magnetic quantum number $m_{a}$.

In an ionization process where the initial state is unpolarized and
the final angular distribution of the free electron is summed, it
is more convenient to consider the process isotropic so the frozen
core potential 
\begin{align}
V_{\textrm{FCA}}^{(a)}(r) & =\frac{Z_{\textrm{eff}}^{(a)}(r)}{r}\nonumber \\
 & =\frac{1}{r}\left(Z-\sum_{b}(2j_{b}+1)v_{b}(r)+v_{a}(r)\right)\,,\label{eq:V_FCA}
\end{align}
has only the radial dependence, and the quantum label $a$ denotes
the ionized shell, and the summation over $b$ is performed to all occupied shells. 
The single-electron potential 
\begin{equation}
v_{a}(r)=\int_{0}^{\infty}dr^{'}\frac{1}{r_{>}}\left(g_{a}(r^{'})^{2}+f_{a}(r^{'})^{2}\right)\,,\label{eq:v_a}
\end{equation}
where $r_{>}$ is the larger one of $r$ and $r^{'}$. The physical
meaning of $V_{\textrm{FCA}}^{(a)}(r)$ is clear: In a mean field
approach where all electrons are treated as independent particles,
the Coulomb potential felt by an ionized $a$-shell electron is the
sum of contributions from the nucleus (of charge $Z$) and all electrons
(the summation $\sum_{b}$ with a degeneracy factor $2j_b+1$) except for itself ($v_{a}$). As Eq.~(\ref{eq:V_FCA}) looks very similar to 
the direct term in a typical Dirac-Hatree-Fock approximation, the potential is also called the 
relativistic Hartree potential (see, e.g., Refs.~\cite{Slater_1951hf,Sampson:1989df}).

We remark that there are other ways of implementing the frozen core approximation.
In some approaches, it is even used to replace the
two-body Coulomb potential, so the self-consistent mean field equation can
be solved more efficiently. When it comes to comparisons of difference
atomic approaches, these details can be important.

\begin{figure*}
	\begin{center}
		
		\begin{tabular}{cc}
			\includegraphics[width=0.4\paperwidth]{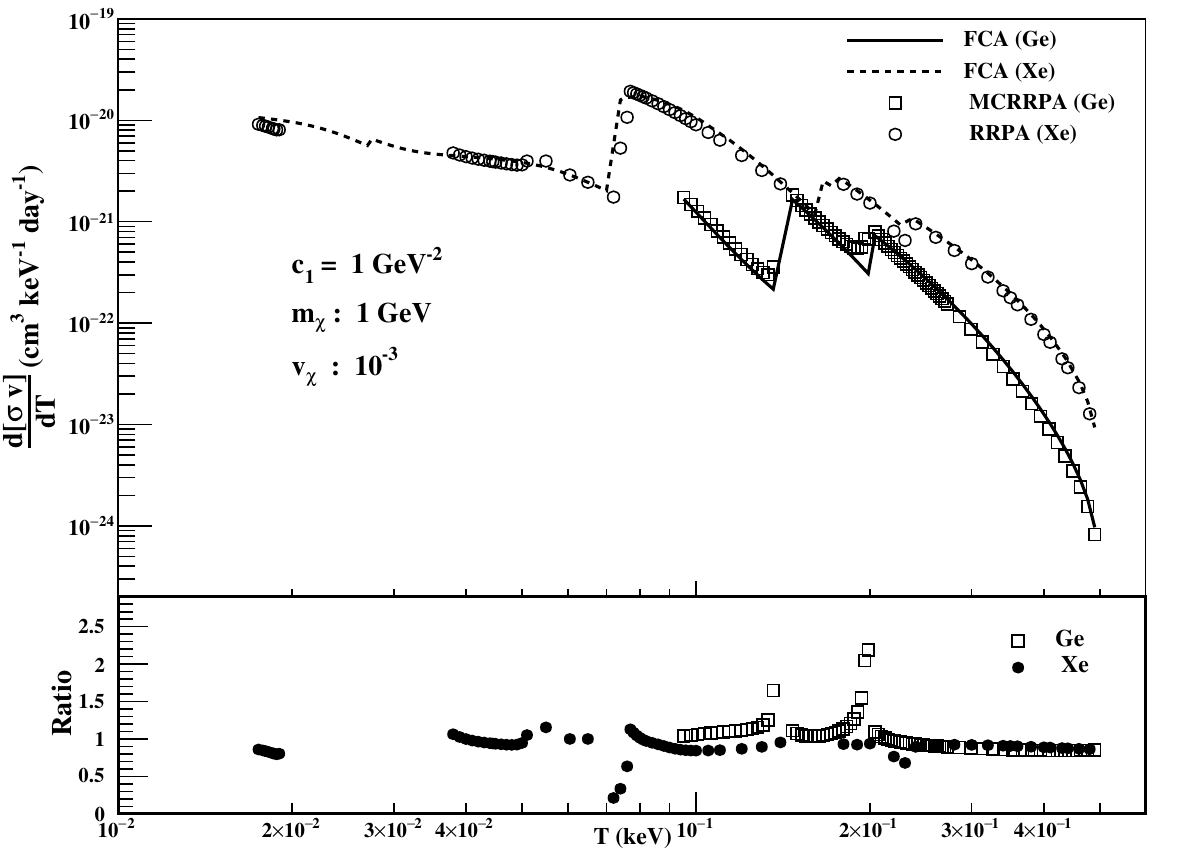} & \includegraphics[width=0.4\paperwidth]{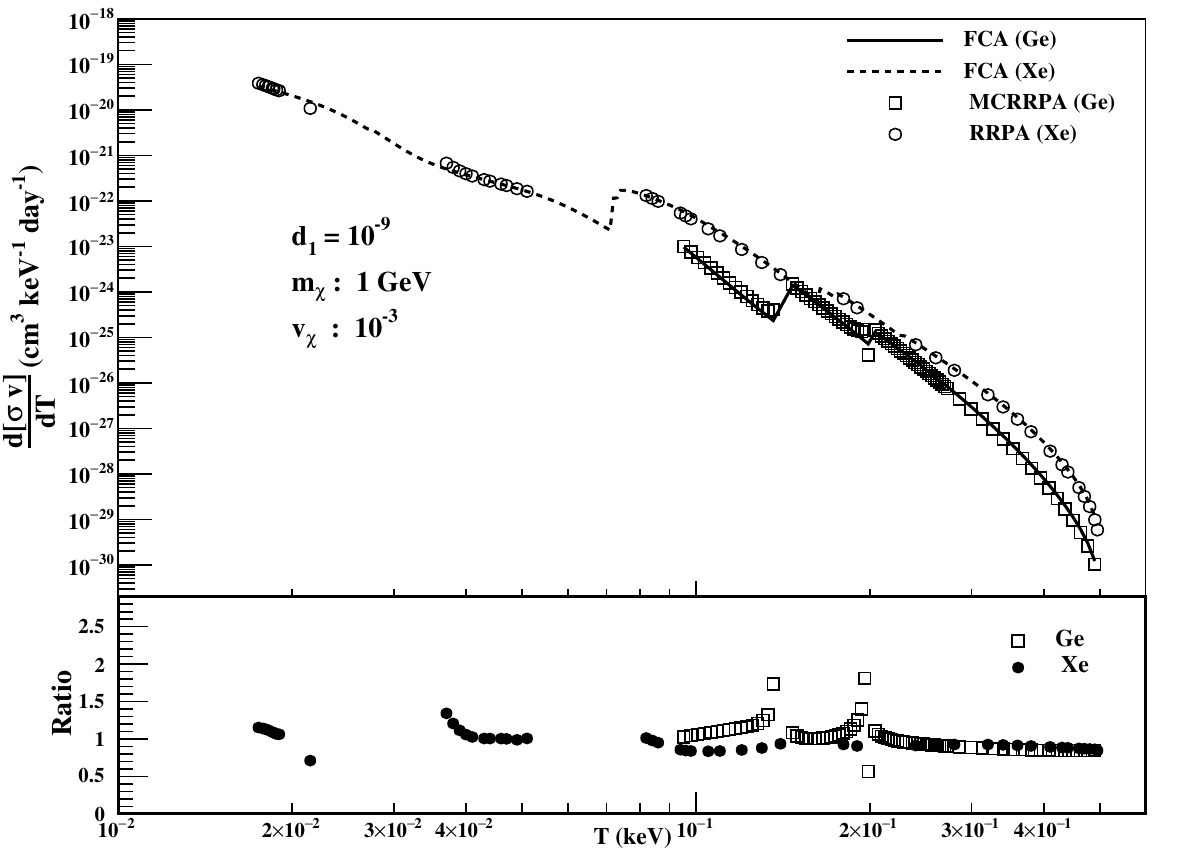}\tabularnewline
		\end{tabular}
		
	\end{center}
	\caption{
		FCA benchmarked by (MC)RRPA. The ratio is for (MC)RRPA/FCA.
		The agreement is within $20\%$ except in the small energy range near the ionization edges.}
	\label{fig::FCAvsRRPA}
\end{figure*}

The procedure we just introduce can be easily implemented in typical
non-relativistic Hartree-Fock schemes, which solve the Schrödinger
equation instead. The only change in formulae above is the wave function
now takes a 1-spinor form, so there is no small component $f_{a}$,
and the quantum label $a=(n_{a},l_{a})$, where $l_{a}$ is the familiar
orbital quantum number.

In Fig.~\ref{fig::FCAvsRRPA}, the FCA results (in lines) are compared
with the (MC)RRPA (in dots). The agreement is generally good at the
level of $20\%$, except for regions close to the ionization edges.

\section{B. Folding with Dark Matter Velocity Spectrum~\label{sec:eta}}

Combining Eqs.~(\ref{eq:dcs-LO} and \ref{eq:vchi_ave}), the computation
of averaged velocity-weighted differential cross section involves
a double integration 
\begin{align}
\frac{d\braket{\sigma v_{\chi}}}{dT}= & \int_{v_{\textrm{min}}}^{v_{\max}}d^{3}v_{\chi}f(\vec{v}_{\chi})v_{\chi}\nonumber \\
 & \times\frac{1}{2\pi v_{\chi}^{2}}\int_{q_{-}}^{q_{+}}dq\, q\left[\left|c_{1}+\frac{d_{1}}{q^{2}}\right|^{2}\right]R(T,q))\,.\label{eq:vdsdT_double}
\end{align}
The integrand has an apparent $v_{\chi}$ dependence which is $v_{\chi}\times v_{\chi}^{-2}=\frac{1}{v_{\chi}}$,
and an implicit one in $q_{\pm}$, which determine the allowed range
of 3-momentum transfer under the given kinematics. Therefore, an
interchange of integration order needs to preserve the phase space
for it to be exact; or at least the part where most strength lies
so it is a good approximation. 

A standard procedure of interchanging integration order  (see, e.g., Refs.~\cite{Kopp:2009et,Essig:2011nj}) is the following:
From energy and momentum conservation, the threshold velocity that
a LDM can instigate an energy transfer $T$ and 3-momentum transfer $q$ is 
\begin{equation}
v_{\chi}\gtrsim\tilde{v}_{\min}=\frac{T}{q}+\frac{q}{2m_{\chi}}\,.\label{eq:v_min}
\end{equation}
Fixing the $q$ integration range by equating $v_{\textrm{min}}=\nicefrac{T}{q_{-}}+\nicefrac{q_{-}}{2m_{\chi}}$ 
and $q_{+}=2m_{\chi}v_{\max}$ so that $q_{\pm}$ no longer have $v_{\chi}$
dependence, then the integration of $v_{\chi}$ can simply be reduced
to a function 
\begin{equation}
\eta(\tilde{v}_{\textrm{min}})=\int_{v_{\min}}^{v_{\max}}d^{3}v_{\chi}f(\vec{v}_{\chi})\frac{1}{v_{\chi}}\Theta(v_{\chi}-\tilde{v}_{\min})\,,\label{eq:eta_func}
\end{equation}
where the step function $\Theta(v_{\chi}-\tilde{v}_{\textrm{min}})$
imposes the velocity requirement. The analytic functional form of
$\eta(\tilde{v}_{\textrm{min}})$ can be found, e.g., in Ref.~\cite{Lewin:1995rx,Savage:2008er}. We have verified within the FCA that the $\eta$ function approach yields the same result as the more cumbersome double integration of Eq.~(\ref{eq:vdsdT_double}) for kinematics relevant in this work. This is contrary to the assertion of Ref.~\cite{Roberts:2019chv}.

\section{C. Comparison of Atomic Approaches to Continuum States~\label{sec:approaches}}

The existing works, including our FCA approach, on DM scattering off
atomic electrons are all based on mean field approaches, either non-relativistic~\cite{Kopp:2009et,Essig:2011nj,Essig:PRL2012,EssigPRD:2017,Agnes:2018ves,Catena:2019gfa} or relativistic~\cite{Roberts:2015lga,Roberts:2016xfw,Roberts:2019chv}.
The first common feature of all is solving the single-electron orbital functions self-consistently from the Hartree-Fock or Dirac-Fock equation. With these orbitals, a good description of the atomic ground state is constructed from one or a linear combination of Slater determinants by filling $Z$ electrons. While relativity
does introduce slight changes of the eigenenergies and eigenfunctions of occupied orbitals, it is fair to say that all approaches have very similar atomic initial states to start with, and it is the atomic
final states that different approaches diverge.

So far, except for the limited data we carried out with (MC)RRPA, the second common feature of all is the ionized electron state is obtained by solving some mean field equation.  Except Refs.~\cite{Roberts:2016xfw,Roberts:2019chv}, it simply takes a Schrödinger or Dirac form with an effective,
isotropic, electrostatic potential 
\begin{equation}
V^{(a)}(r)=\frac{Z^{(a)}(r)}{r}\,,\label{eq:V_(a)}
\end{equation}
where $a$ is the quantum label of the ionized atomic shell. A few prescriptions
of the effective charge $Z^{(a)}(r)$ are listed in the following. 
\begin{itemize}
\item Plane wave approximation (PWA)~\cite{Kopp:2009et,Essig:2011nj}:
This is done by simply taking $Z^{(a)}=0$, i.e., the ionized electron
is completely free. Calculation-wise, the transition matrix element
is simplified to the Fourier transform of the bound state wave function.
However, unless the scattering energy is much higher than the atomic
scale, this assumption can hardly be justified.
\item Hydrogen-like approximation (HLA)~\cite{Essig:2011nj,Agnes:2018ves,Catena:2019gfa}:
This approximation assumes the ionized electron behaves like a hydrogen-like
electron. The effective charge is simply a unscreened point source
at the atomic center with its magnitude fixed by the orbital binding
energy $E_{B}$: 
\begin{equation}
Z_{\textrm{HLA}}^{(a)}(r)=n_{a}\sqrt{\frac{-E_{B}^{(a)}}{\textrm{Ry}}}\,,\label{eq:Z_HLA}
\end{equation}
where $n_{a}$ is the principle quantum number (in the NR case, $n_{a}$
is an integer; in the relativistic case, there is some correction);
and $\textrm{Ry}=13.6\,\textrm{eV}$. While this approximation is
tuned to reproduce the correct binding energy, this point charge picture
is not realistic, either.
\item Frozen core approximation (FCA):
This approximation has been explained in Appendix A. We additionally
comment here that the resulting effective charge has the correct asymptotic
behaviors: i.e. $Z_{\textrm{FCA}}^{(a)}(r\rightarrow r_{N})\rightarrow Z$
where $r_{N}$ is the nuclear surface and $Z_{\textrm{FCA}}^{(a)}(r\rightarrow\infty)\rightarrow1$.
Therefore, it is more realistic than PWA and HLA.
\end{itemize}

\begin{figure}
	\begin{center}
		
		\includegraphics[scale=0.08]{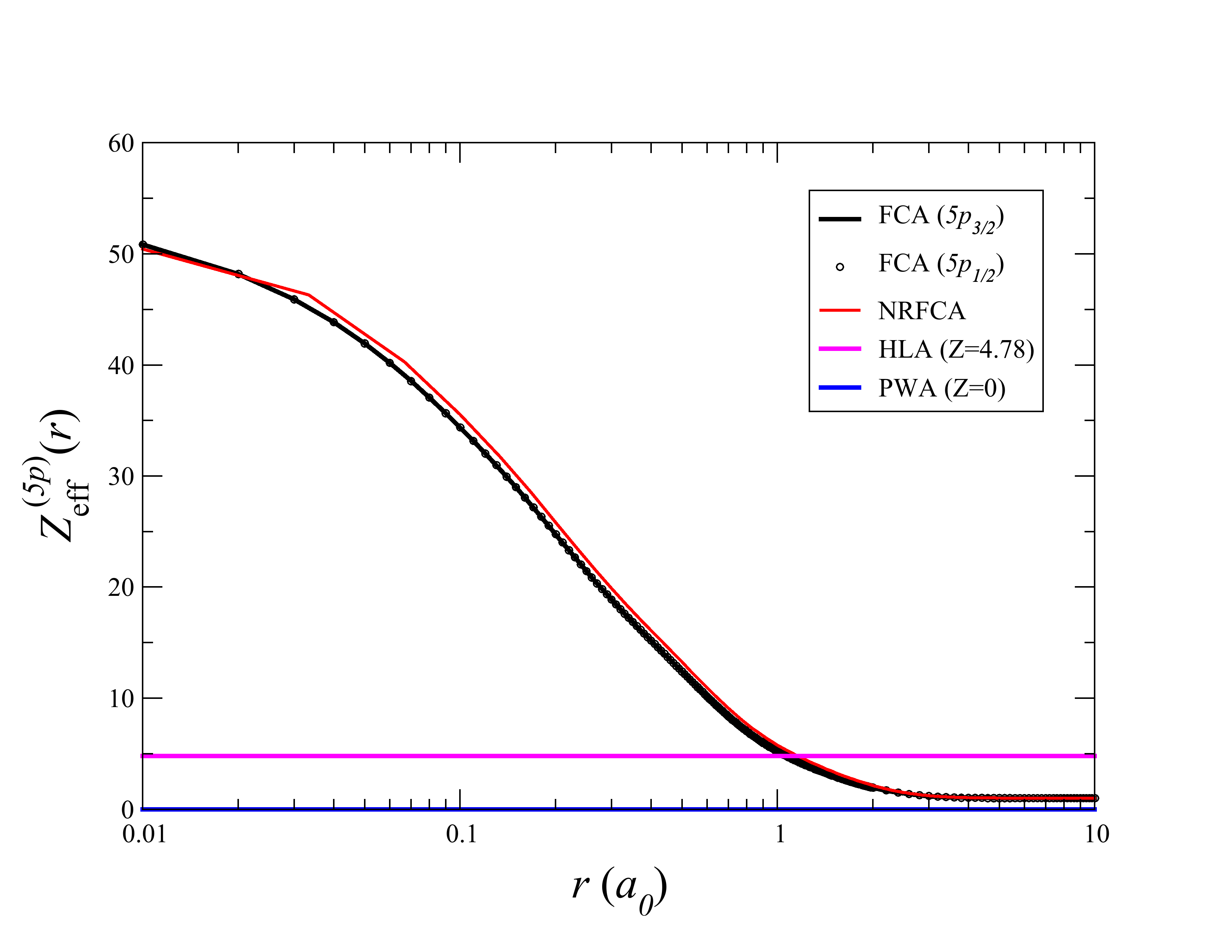}
		
	\end{center}\caption{The effective charge $Z_{\textrm{eff}}^{(5p)}(r)$ felt by the electron
		ionized from a $5p$ orbital derived from the approaches of FCA, NRFCA,
		HLA, and PWA. Note that the difference between relativistic $5p_{3/2}$
		and $5p_{1/2}$ are barely visible.}
	\label{fig::Zeff}
\end{figure}

In Fig.~\ref{fig::Zeff}, we use the xenon $5p$ shell(s) as an example
to illustrate the differences in $Z_{\textrm{eff}}^{(5p)}$.
Obviously, when the 3-momentum transfer $q$ associated in DM-atom scattering is $\gtrsim\textrm{keV}$, i.e., the inverse of atomic size, the scattering amplitude depends 
sensitively on the effective charge distribution. On the other hand, for much
smaller or larger $q$, atomic physics still plays an important role
in getting reliable results. For the former, the correct asymptotic
requirement $Z_{\textrm{FCA}}^{(a)}(r\rightarrow\infty)\rightarrow1$
is still crucial (otherwise phase shifts would not be correctly inferred).
For the latter, the relativistic and nuclear finite-size effect are
both significant, as already pointed out in Refs.~\cite{Roberts:2015lga,Roberts:2016xfw}. 

\bibliographystyle{apsrev4-1}
\bibliography{dmes_SI}

\end{document}